\documentclass[a4paper,11pt]{article}

\usepackage{amsmath,amsthm,amsfonts}

\usepackage{cite}
\usepackage{mathptmx,times}

\newcommand{\HH}{\mathcal{H}}

\newcommand{\CC}{\mathbb{C}}
\newcommand{\RR}{\mathbb{R}}
\newcommand{\ZZ}{\mathbb{Z}}
\newcommand{\NN}{\mathbb{N}}
\newcommand{\GG}{\mathcal{G}}

\newcommand{\dom}{\mathop{\mathrm{dom}}}
\newcommand{\diag}{\mathop{\mathrm{diag}}}

\newcommand{\spec}{\mathop{\mathrm{spec}}}

\newcommand{\tr}{\mathop{\mathrm{tr}}}

\newtheorem{theorem}{Theorem}
\newtheorem{prop}[theorem]{Proposition}
\newtheorem{corol}[theorem]{Corollary}

\theoremstyle{definition}

\newtheorem{rem}[theorem]{\bf Remark}
\newtheorem{exm}[theorem]{\bf Example}

\begin{document}

\title{Cantor and band spectra for periodic quantum graphs with~magnetic fields\thanks{The work was supported by the Deutsche Forschungsgemeinschaft, the Sonderforschungsbereich ``Raum, Zeit, Materie'' (SFB 647),
and the International Bureau of BMBF at the German Aerospace Center
(IB DLR, cooperation Germany -- New Zealand NZL 05/001)}}

\author{Jochen Br\"uning\dag\and Vladimir Geyler\ddag\and Konstantin Pankrashkin\dag\S\\[\bigskipamount]
\dag Institut f\"ur Mathematik, Humboldt-Universit\"at zu Berlin,\\ Rudower Chaussee 25, 12489 Berlin, Germany\\
\ddag Mathematical Faculty, Mordovian State University,\\ 430000 Saransk, Russia\\
\S Corresponding author, E-mail const@mathematik.hu-berlin.de\\
Phone +49 30 2093 2352, Fax +49 30 2093 2727}

\date{}

\maketitle

\begin{abstract}
We provide an exhaustive spectral analysis of the two-dimensional periodic square
graph lattice with a magnetic field. We show that the spectrum
consists of the Dirichlet eigenvalues of the edges
and of the preimage of the spectrum of a certain discrete operator
under the discriminant (Lyapunov function) of a suitable Kronig-Penney Hamiltonian.
In particular, between any two Dirichlet
eigenvalues the spectrum is a Cantor set  for an irrational flux, and is absolutely continuous
and has a band structure for a rational flux. The Dirichlet eigenvalues
can be isolated or embedded, subject to the choice
of parameters. Conditions for both possibilities are given.
We show that generically there are infinitely many gaps in the spectrum,
and the Bethe-Sommerfeld conjecture fails in this case.
\end{abstract}

\section*{Introduction}

The Hamiltonian $H$ of a two-dimensional Bloch electron in a
uniform magnetic field $B$ has a highly non-trivial spectral and
topological structure depending on the ratio of the area $\sigma$
of the elementary cell of the lattice in question and the squared
magnetic length $\ell_M^2=\hbar c/|eB|$ (here $e$ is the electron
charge, $c$ is the light velocity, $\hbar$ is the Planck
constant). More precisely, denote by $\theta$ the number of the
magnetic flux quanta through the elementary cell:
$\theta=\sigma/2\pi\ell_M^2$. If $\theta$ is a rational number,
then the spectrum of $H$ has a band structure and for $\theta\ne0$
each vector bundle of the magnetic Bloch functions corresponding
to a completely filled Landau levels is non-trivial~\cite{nov}
(for a non-zero integer $\theta$ the Chern number of this bundle
is exactly the values of the quantized Hall conductance in units
$e^2/2\pi\hbar$~\cite{TKNN,Av,Pr}). The most likely conjecture is
that the at irrational values of $\theta$, the spectrum of $H$ has
a Cantor structure. This conjecture was recently proved in the
case of the tight-binding model for the magnetic Bloch
electron~\cite{AK,10mart,pu}; in this case $H$ is reduced to the
Harper operator in the discrete Hilbert space $l^2(\mathbb{Z)}$.
Due to the Langbein duality~\cite{lb}, the same is true for the
Hamiltonian of the nearly-free-electron approximation. As a result
of this Cantor properties, the diagram describing the dependence
of the spectrum on the flux $\theta$, so-called Hofstadter
%diagram,
butterfly, has a remarkable fractal structure, see
e.g.~\cite{hof,hk,kr}.

A few is known for the magnetic Schr\"odinger operator $H$ in the
space $L^2(\mathbb{R}^2)$,
\begin{equation}
\label{Ham} H=\frac{1}{2m}\left({\bf p}-\frac{e}{c}{\bf
A}\right)^2+V(x,y)\,,
\end{equation}
where ${\bf A}$ is the vector potential of $B$ and $V$ is a
potential which is periodic with respect to the considered
lattice. It is proven in this case, that $H$ has a piece of the
Cantor spectrum near the bottom of the spectrum for a restricted
class of potentials $V$~\cite{HS}.

In this connection, the quantum network models (also called the
quantum graph models) have attracted considerable interest last
time. These models combine some essential features of both
discrete and continuous models mentioned above. On the one hand,
the Hamiltonian of a magnetic network model has infinitely many
magnetic bands of different shape. On the other, the
time-independent Schr\"odinger equation for this Hamiltonian can
be reduced to a discrete equation. S.~Alexander was the first who
performed this reduction~\cite{alx} in the framework of the 
percolation approach to the effect of disorder
on superconductivity proposed
by P.~G.~de~Gennes~\cite{dGe}. A very short and elegant derivation
of the Schr\"odinger equation for a periodic quantum graph with a
uniform magnetic field and a constant potential on each edge of
the graph is given in~\cite{khm}. On the mathematical level of
rigor the relation between solutions of the Schr\"odinger equation
for $H$ on quantum graphs $\Gamma$ and those for a Jacoby matrix
$J(H)$ on the corresponding combinatorial graphs was established by
P.~Exner~\cite{exdual}. Nevertheless, the main theorem
from~\cite{exdual} is applicable only to the finding of
eigenvalues of $H$ distinct from the Dirichlet eigenvalues on the
edges of $\Gamma$. As to the points of the continuous spectrum,
the main result of~\cite{exdual} allows an exhaustive analysis
only in the case when the direct and inverse Schnol-type theorems
are known for both $H$ and $J(H)$.

It is worthy to note that quantum networks are not only a
mathematical tool to get simplified models of various quantum
systems, but in many cases experimental devices really have a
shape of planar graphs such that the width of the sides is much
smaller than the parameters of the dimension of length which
characterizes the quantity in question, e.g., much smaller that
the magnetic length, the Fermi wave length, the scattering length
etc~\cite{p1,p2,p3}. In these cases the quantum graph models are
the most adequate ones for simulating spectral, scattering, and
transport properties of these devices.

Here we propose an alternative approach to the spectral analysis
of quantum graph Hamiltonians based on boundary triples,
Dirichlet-to-Neumann maps, and the Krein technique of self-adjoint extensions.
Such a machinery works effectively in many other problems connected with explicitly
solvable models~\cite{AGHH,pavlov}.
In the case of square network lattices with a periodic magnetic field
(including a uniform one), an arbitrary $L^2$-potential on edges
and $\delta$-like boundary conditions at the vertices (including
the Kirchhoff boundary conditions), we perform an exhaustive
spectral analysis of the network Hamiltonian $H$. It is proved
that the spectrum always contains Dirichlet eigenvalues of the
edges as infinitely degenerate eigenvalues of $H$. The rest part
of the spectrum is absolutely continuous and has a band structure,
if $\theta$ is a rational number, and is the union of countably
many Cantor sets placed between Dirichlet
eigenvalues, otherwise. Moreover, this part is the preimage of the
spectrum of the corresponding lattice Hamiltonian $J(H)$ with
respect to a many-sheeted analytic function which is just the
discriminant of a generalized Kronig--Penney operator (i.e. the
Sturm--Liouville operator with a Kronig--Penney potential). The
 eigenvalues are isolated if the magnetic
flux is non-integer, while for integer magnetic fluxes it depends
on the electric potential and on the coupling at the nodes.

\section{Magnetic Schr\"odinger operator on the periodic metric graph}\label{sec2}
Consider a planar square graph lattice whose nodes are the points
$K_{m,n}:=(m l,n l)$, $(m,n)\in\ZZ^2$, where $l>0$ is the length of each edge.
Two nodes $K_{m,n}$ and $K_{p,q}$ are connected by an edge
iff $|m-p|+|n-q|=1$. We denote the edge between $K_{m,n}$ and $K_{m+1,n}$
by $E_{m,n,r}$ (right), and between  $K_{m,n}$ and $K_{m,n+1}$ by $E_{m,n,u}$ (up).
Each edge $E_{m,n,r/u}$ will be considered as segment $[0,l]$
so that $0$ is identified in the both cases with $K_{m,n}$, and $l$ is identified
with $K_{m+1,n}$ for $E_{m,n,r}$ and $K_{m,n+1}$ for $E_{m,n,u}$, respectively.
The phase space of the lattice is
\[
\HH=\bigoplus_{(m,n)\in\ZZ}\big(\HH_{m,n,r}\oplus\HH_{m,n,u}\big),\quad
\HH_{m,n,r/u}=L^2[0,l].
\]
The elements of $\HH$ will be denoted as $f=(f_{m,n,r},f_{m,n,u})$,
$f_{m,n,r/u}\in\HH_{m,n,r/u}$.
On each edge consider the same electric potential $V\in L^2[0,l]$.

We assume that the lattice is subjected to an external
periodic magnetic field orthogonal to the plane,
$B(x)=\big(0,0,b(x)\big)$, $b\in C(\RR^2)$,
$b(x_1,x_2)=b(x_1+l,x_2)=b(x_1,x_2+l)$ for all $x=(x_1,x_2)\in\RR^2$.
Denote
\[
\xi=\frac{1}{2\pi l^2}\,\int_F b(x)dx,\quad
\]
where $F$ is the fundamental domain of the lattice (i.e. the
square spanned by $E_{0,0,r}$ and $E_{0,0,u}$) and represent
$B(x)=B_0+B_p(x)$, $B_0=(0,0,2\pi\xi)$. The corresponding magnetic
vector potential in the symmetric gauge can be
%represented
written as
\[
A(x_1,x_2,x_3)=\frac{1}{2} \,B_0\times (x_1,x_2,x_3)+\big(A_1(x_1,x_2),A_2(x_1,x_2),0\big)
\]
with $A_j(x_1,x_2)=A_j(x_1+l,x_2)=A_j(x_1,x_2+l)$, $j=1,2$.
The presence of the magnetic field leads to non-trivial
magnetic potentials on the edges, which are the projections of $A(x)$ on the corresponding
directions. The magnetic potentials $A_{m,n,r/u}$ on $E_{m,n,r/u}$
are:
\begin{align*}
A_{m,n,r}(t)&=\big\langle A\big((ml,nl,0)+(1,0,0)t\big),(1,0,0)\big\rangle%\\
\equiv-\pi nl\xi+a_r(t),\\
A_{m,n,u}(t)&=\big\langle A\big((ml,nl,0)+(0,1,0)t\big),(0,1,0)\big\rangle
\equiv \pi ml\xi+a_u(t),\\
a_r(t)&:=A_1(t,0),\\
a_u(t)&:=A_2(0,t),\quad t\in[0,l].
\end{align*}
On each of the edges $E_{m,n,r/u}$ we consider the operator
\[
L_{m,n,r/u}=\Big(-i\frac{d}{dt}-A_{m,n,r/u}\Big)^2+V,\quad
\]
with the domain $H^2[0,l]$. The direct sum of these operators over all edges
is not self-adjoint, and in order to obtain a self-adjoint operator on the whole lattice
it is necessary to introduce boundary conditions at each node.
The most general boundary conditions involve a number of parameters
and can be found, for example, in~\cite{KSM}. We restrict
ourselves by considering the so-called magnetic $\delta$-like
interaction at $K_{m,n}$,
\begin{equation}
        \label{eq-BC1}
\begin{gathered}
f_{m,n,r}(0)=\frac{1}{\beta}\,f_{m,n,u}(0)=f_{m-1,n,r}(l)=\frac{1}{\beta}\,f_{m,n-1,u}(l)=:f_{m,n},\\
\big(\frac{d}{dt}-iA_{m,n,r}\big)f_{m,n,r}(0)+\beta\big(\frac{d}{dt}-iA_{m,n,u}\big)f_{m,n,u}(0)\\
-\big(\frac{d}{dt}-iA_{m-1,n,r}\big)f_{m-1,n,r}(l)-\beta\big(\frac{d}{dt}-iA_{m,n-1,u}\big)f_{m,n-1,u}(l)=
\alpha f_{m,n},\\m,n\in\ZZ,
\end{gathered}
\end{equation}
where $\alpha\in\RR$, $\beta\in\RR\setminus\{0\}$. These
quantities have the following physical meaning. The parameter
$\alpha\ne0$ is the coupling constant of a $\delta$-like potential
at each node. Introducing the parameter $\beta$ can be
treated as considering a more general form of the Hamiltonian $H$:
\[
H=\sum_{jk}\frac{1}{2m_{jk}}\left(p_j-\frac{e}{c}A_j\right)
\left(p_k-\frac{e}{c}A_k\right)+V(x,y)\,,
\]
where $m_{jk}$ is the  effective mass tensor and $\beta$ is the
corresponding anisotropy coefficient (the ratio of the eigenvalues
of the symmetric matrix $(m_{jk}))$. In particular, if $\beta=1$,
one obtains $H$ in the form \eqref{Ham}; if in addition
$\alpha=0$, we get the magnetic Kirchhoff coupling. This class of
boundary conditions covers main couplings used in the physics
literature. The self-adjoint operator obtained in this way we
denote by $L$.

\begin{rem}\label{rem-mtg}
It is worthy to note that $L$ is invariant with respect to the so-called
magnetic translation group $\mathbb{G}_M$~\cite{Zak}.
In our case this group is generated by the magnetic shift operators
$\tau_r$ and $\tau_u$,
\begin{align*}
\tau_r f_{m,n,r/u}(t)&=\exp\bigg[\pi i\theta\Big(n+\delta_{u,r/u}\frac{\,t\,}{l}\,\Big)\bigg]f_{m-1,n,r/u}(t),\\
\tau_u f_{m,n,r/u}(t)&=\exp\bigg[-\pi i\theta\Big(m+\delta_{r,r/u}\frac{\,t\,}{l}\,\Big)\bigg]f_{m,n-1,r/u}(t).
\end{align*}
The properties of this group depend drastically
on the arithmetic properties of $\theta$ \cite{B}.
In particular, if $\theta$ is irrational, then $\mathbb{G}_M$
has only infinite-dimensional irreducible representations which are
trivial on the center of $\mathbb{G}_M$. Therefore, for any irrational $\theta$
each point of $\spec L$ is infinitely degenerate.
\end{rem}

\section{Gauge transformations}
To study the spectral properties of $L$ it is useful to use the gauge transformation
$(f_{m,n,r},f_{m,n,u})=
(U_{m,n,r}\varphi_{m,n,r},U_{m,n,u}\varphi_{m,n,u})$
given by
\begin{align*}
f_{m,n,r/u}(t)&=\exp\Big(
i\int_0^t A_{m,n,r/u}ds
\Big) \varphi_{m,n,r/u}(t)=:U_{m,n,r/u}\varphi_{m,n,r/u}(t),\\
U_{m,n,r}\varphi_{m,n,r}(t)&= e^{-i\pi n l\xi t+b_r(t)}\varphi_{m,n,r}(t),\\
U_{m,n,u}\varphi_{m,n,u}(t)&=e^{i\pi m l\xi t+b_u(t)}\varphi_{m,n,u}(t),\\
b_{r/u}(t)&:=\int_0^t a_{r/u}(s)\,ds.
\end{align*}
There holds
$U_{m,n,r/u}^{-1}L_{m,n,r/u}U_{m,n,r/u}=-\dfrac{d^2}{dt^2}+V$,
and the boundary conditions \eqref{eq-BC1} for $\varphi=(\varphi_{m,n,r},\varphi_{m,n,u})\in U^{-1}(\dom L)$ take the form
\begin{subequations}
        \label{eq-BC-ini}
\begin{align}
\Tilde \varphi'_{m,n}&=\alpha\Tilde \varphi_{m,n},\\
\intertext{where}
\Tilde \varphi_{m,n}&:=
\varphi_{m,n,r}(0)=\frac{1}{\beta}\,\varphi_{m,n,u}(0)\notag\\
&=e^{-i\pi n\theta+\kappa_r}\varphi_{m-1,n,r}(l)
=\frac{1}{\beta}\,e^{i\pi m\theta+\kappa_u}\varphi_{m,n-1,u}(l),\\
\Tilde \varphi'_{m,n}&:=\varphi'_{m,n,r}(0)+\beta\varphi'_{m,n,u}(0)\notag \\
&\quad{}-e^{-i\pi n \theta+\kappa_r}\varphi'_{m-1,n,r}(l)- \beta e^{i\pi m\theta+\kappa_u}\varphi'_{m,n-1,u}(l),\\
\kappa_{r/u}&:=b_{r/u}(l),\qquad m,n\in\ZZ,\notag
\end{align}
\end{subequations}
and $\theta:=\xi l^2$ is the number of flux quanta through the elementary cell.
Therefore, the operator $\Tilde L=U^{-1} L U$ acts on each edge as $\varphi_{m,n,r/u}\mapsto -\varphi''_{m,n,r/u}+V\varphi_{m,n,r/u}$
on functions $\varphi$ satisfying~\eqref{eq-BC-ini}, and its spectrum coincides with the spectrum of $L$.

To simplify subsequent calculations we apply another gauge transformation,
$\varphi_{m,n,r/u}=\exp\big(i(m\kappa_r+n\kappa_u)\big)\phi_{m,n,r/u}$.
In this way we arrive et an operator acting on each edge
as $\phi_{m,n,r/u}\mapsto -\phi''_{m,n,r/u}+V\phi_{m,n,r/u}$
on functions $\phi=(\phi_{m,n,r},\phi_{m,n,u})$, $\phi_{m,n,r/u}\in H^2[0,l]$, satisfying
\begin{subequations}
        \label{eq-BC2}
\begin{align}
        \label{eq-BC2a}
\phi'_{m,n}&=\alpha\phi_{m,n},\qquad m,n,\in\ZZ,\\
        \label{eq-BC2a1}
\phi_{m,n}&:=
\phi_{m,n,r}(0)=\frac{1}{\beta}\,\phi_{m,n,u}(0)=e^{-i\pi n\theta}\phi_{m-1,n,r}(l)=\frac{1}{\beta}\,e^{i\pi m\theta}\phi_{m,n-1,u}(l),\\
        \label{eq-BC2b}
\phi'_{m,n}&:=\phi'_{m,n,r}(0)+\beta\phi'_{m,n,u}(0)-e^{-i\pi n \theta}\phi'_{m-1,n,r}(l)- \beta e^{i\pi m\theta}\phi'_{m,n-1,u}(l),
\end{align}
\end{subequations}
This operator, which we denote by $\Lambda$, is unitarily equivalent
to the initial magnetic Hamiltonian. In what follows we work
mostly with this new operator.

At this point we emphasize two important circumstances. First, for
usual magnetic Schr\"odinger operators in $L^2(\RR^2)$ the
spectral analysis for non-zero but periodic magnetic vector
potentials (i.e. with the zero flux per cell) essentially differs
from that for the Schr\"odinger operators without magnetic field.
Even the proof of the absolute continuity of the spectrum is
non-trivial~\cite{BS,Sob}. In our case, the operator on the graph
with a periodic magnetic vector potential appears to be unitarily
equivalent to the operator without magnetic field. Second, for the
usual magnetic Schr\"odinger operators the bottom of the spectrum
grows infinitely as the flux becomes infinitely large. In our
situation, the spectrum is 1-periodic with respect to the magnetic
flux $\theta$, as changing $\theta$ by $\theta+1$
in~\eqref{eq-BC2} obviously can compensated by a unitary
transformation.
%In corresponding physical quantities such
%periodicity leads to the so-called Aharonov--Bohm oscillations.
Such periodicity leads to the so-called Aharonov--Bohm
oscillations in the corresponding physical quantities.

\begin{prop}
The operator $\Lambda$ is semibounded below.
\end{prop}

\begin{proof}
Let $\phi\in\dom \Lambda$. Using the integration by parts
and changing suitably the summation order  one obtains
\begin{multline*}
\langle \phi,\Lambda\phi\rangle=\sum_{m,n}\big(
\langle \phi_{m,n,r},-\phi''_{m,n,r}+V\phi_{m,n,r}\rangle+
\langle \phi_{m,n,u},-\phi''_{m,n,u}+V\phi_{m,n,u}\rangle\big)\\
{}=\sum_{m,n}\Big(
\overline{\phi_{m,n,r}(0)}\phi'_{m,n,r}(0)-\overline{\phi_{m,n,r}(l)}\phi'_{m,n,r}(l)%\\
{}+\overline{\phi_{m,n,u}(0)}\phi'_{m,n,u}(0)-\overline{\phi_{m,n,u}(l)}\phi'_{m,n,u}(l)\\
{}+\int_0^l \big(|\phi'_{m,n,r}|^2+V|\phi_{m,n,r}|^2\big)dx+\int_0^l \big(|\phi'_{m,n,u}|^2+V|\phi_{m,n,u}|^2\big)dx
\Big)\\
{}=\sum_{m,n}\Big\{
\int_0^l \big(|\phi'_{m,n,r}|^2+V|\phi_{m,n,r}|^2\big)dx+\int_0^l \big(|\phi'_{m,n,u}|^2+V|\phi_{m,n,u}|^2\big)dx\\
{}+\overline{\phi_{m,n}}\,\big(\phi'_{m,n,r}(0)+\beta\phi'_{m,n,u}(0)-e^{-i\pi n\theta}\phi'_{m-1,n,r}(l)
-\beta e^{i\pi m\theta}\phi'_{m,n-1,u}(l)\big)\Big\}
\end{multline*}
\begin{multline*}
{}=\sum_{m,n}\Big\{
\int_0^l \big(|\phi'_{m,n,r}|^2+V|\phi_{m,n,r}|^2\big)dx+\int_0^l \big(|\phi'_{m,n,u}|^2+V|\phi_{m,n,u}|^2\big)dx+\overline{\phi_{m,n}}\,\phi'_{m,n}
\Big\}\\
{}=\sum_{m,n}\Big\{
\int_0^l \big(|\phi'_{m,n,r}|^2+V|\phi_{m,n,r}|^2\big)dx+\int_0^l \big(|\phi'_{m,n,u}|^2+V|\phi_{m,n,u}|^2\big)dx+\alpha|\phi_{m,n}|^2\Big\}.
\end{multline*}
Now choose $c\in (0,1)$ and $C\in \RR$ with
\[
|\alpha||h(0)|^2\le \int_0^l\big(c|h'|^2+(V+C)|h|^2\big)dx \text{ for all } h\in H^1[0,l]
\]
(the existence of such constants follows from the Sobolev inequality), then
\[
|\alpha| |\phi_{m,n}|^2\equiv|\alpha||\phi_{m,n,r}(0)|^2\le \int_0^l\big(c|\phi'_{m,n,r}|^2+(V+C)|\phi_{m,n,r}|^2\big)dx,
\]
and
\[
\langle \phi,(\Lambda+C)\phi\rangle\ge
\sum_{m,n}\,\int_0^l \Big\{(1-c)|\phi'_{m,n,r}|^2+
|\phi'_{m,n,u}|^2+(V+C)|\phi_{m,n,u}|^2\Big\}\,dx\ge0.
\qedhere
\]
\end{proof}

\section{Boundary triples}

Here we describe briefly the technique of abstract self-adjoint boundary value problems
with the help
of boundary triples. For more detailed discussion we refer to~\cite{dm}.

Let $S$ be a closed linear operator in a Hilbert space $\HH$ with the domain $\dom S$.
Assume that there exists
an auxiliary Hilbert space $\GG$ and two linear maps $\Gamma,\Gamma':\dom S\to \GG$ such that
\begin{itemize}
\item for any $f,g\in\dom S$ there holds
$\langle f,Sg\rangle-\langle Sf,g\rangle=\langle  \Gamma f,\Gamma' g\rangle-\langle\Gamma'f,\Gamma g\rangle$,
\item the map $(\Gamma,\Gamma'):\dom S\to\GG\oplus\GG$ is surjective,
\item the set $\ker\,\Gamma \cap\ker\,\Gamma'$ is dense in $\HH$.
\end{itemize}
The triple $(\GG,\Gamma, \Gamma')$ with the above properties
is called a \emph{boundary triple} for $S$.

\begin{exm}
Let us describe one important example of boundary triple.
Let $V\in L^2[0,l]$ be a real-valued function. In $\HH=L^2[0,l]$
consider the operator
\begin{equation}
             \label{eq-os}
S=-\dfrac{d^2}{dt^2}+V,\quad \dom S=H^2[0,l],
\end{equation}
then one can set
\begin{equation}
             \label{eq-gos}
\GG=\CC^2,\quad
\Gamma f=\begin{pmatrix}
            f(0)\\
            f(l)
         \end{pmatrix},
\quad
\Gamma' f=\begin{pmatrix}
           f'(0)\\
           -f'(l)
          \end{pmatrix}.
\end{equation}
\end{exm}

\bigskip

If an operator $S$ has a boundary triple, then it has self-adjoint restrictions
(see proof of theorem~3.1.6 in \cite{gg}), hence $S^*$ is a
symmetric operator. For example, if $T$ is a self-adjoint operator
in $\GG$, then the restriction of $S$ to elements $f$ satisfying
abstract boundary conditions $\Gamma' f=T\Gamma f$ is a
self-adjoint operator in $\HH$, which we denote by $H_T$. Another
example is the operator $H$ corresponding to the boundary
conditions $\Gamma f=0$. One can relate the resolvents of $H$ and
$H_T$ as well as their spectral properties by the  Krein resolvent
formula, which is our most important tool in this paper.

Let $z\notin\spec H$. For $g \in \GG$ denote by $\gamma(z) g$ the unique solution to
the abstract boundary value problem
$(S-z)f=0$ with $\Gamma f=g$ (the solution exists due to the above conditions for $\Gamma$
and $\Gamma'$).
Clearly, $\gamma(z)$ is a linear map from $\GG$ to $\HH$.
Denote also by $Q(z)$ the operator on $\GG$ given by $Q(z)g=\Gamma' \gamma(z)g$;
this map is called the Krein function. The operator-valued functions $\gamma$ and $Q$ are analytic outside $\spec H$. Moreover, $Q(z)$ is self-adjoint for real $z$.

\begin{prop}\label{th-krein}
\begin{itemize}
\item[(A)] {\rm(Proposition 2 in~\cite{dm})} For $z\notin \spec
H\cup\spec H_T$ the operator $Q(z)-T$ acting on $\GG$ has a
bounded inverse defined everywhere and the Krein resolvent formula
holds,
\[
(H-z)^{-1}-(H_T-z)^{-1}=\gamma(z)\big(Q(z)-T\big)^{-1}\gamma^*(\Bar z).
\]
\item[(B)] The set $\spec H_T\setminus\spec H$ consists exactly of
real numbers $z$ such that $0\in\spec \big(Q(z)-T\big)$.

\item[(C)] {\rm (Theorem 1 in~\cite{gm})} Let $z\in \spec
H_T\setminus\spec H$, then $z$ is an eigenvalue of $H_T$ if and
only if $0$ is an eigenvalue of $Q(z)-T$, and in this case
$\gamma(z)$ is an isomorphism of the corresponding eigensubspaces.

%
%\item[(D)] {\rm (Lemma~3.3.1 in~\cite{gg})}  $\Gamma$ and
%$\Gamma'$ are continuous linear mappings from $\dom S$ endowed
%with the graph norm to $\GG$.
\end{itemize}
\end{prop}
This statement is especially useful if the spectrum of $H$ is a
discrete set and the spectrum of $H_T$ is expected to have a
positive measure, because one can describe the most part of the
spectrum of $H_T$ in terms of $Q(z)-T$.

\begin{exm}\label{ex2}
Consider the example given by \eqref{eq-os}
and \eqref{eq-gos}. The corresponding Krein function $s(x)$
can be obtained as follows. The restriction $D$ of $S$ given by $\Gamma f=0$ is
\begin{equation}
     \label{eq-od}
Df=-f''+Vf,\quad \dom D=\{f\in H^2[0,l]:\,f(0)=f(l)=0\}
\end{equation}
In what follows we denote the eigenvalues of $D$ by $\mu_k$, $k=0,1,2\dots,$ $\mu_0<\mu_1<\mu_2<\dots$.

Let two functions $u_1,u_2\in H^2[0,l]$ satisfy
\begin{equation}
       \label{eq-uu12}
u_1,\,u_2\in\ker(S-z),\quad
\begin{aligned}
%-u''_1(x;z)+\big(V(x)-z\big)u_1(x;z)&=0, &
u_1(0;z)&=0, & u'_1(0;z)&=1,\\
%-u''_2(x;z)+\big(V(x)-z\big)u_2(x;z)&=0, &
u_2(0;z)&=1, & u'_2(0;z)&=0.
\end{aligned}
\end{equation}
Clearly, for their Wronskian one has
$w(z)=u'_1(x;z)u_2(x;z)-u_1(x;z)u'_2(x;z)\equiv 1$ Both $u_1$,
$u_2$ as well as their derivatives with respect to $x$ are entire
functions of $z$.

Let $z\notin\spec D$, then any function $f$ with $-f''+Vf=zf$ can be written as
\[
f(x;z)=\frac{f(l)-f(0) u_2(l;z)}{u_1(l;z)}\,u_1(x;z)+f(0)u_2(x;z),
\]
and the calculation of $f'(0)$ and $-f'(l)$ gives
\begin{equation}
               \label{eq-rszA}
s(z)=\frac{1}{u_1(l;z)}\,\begin{pmatrix}
-u_2(l;z) & 1\\
w(z) & -u'_1(l;z)
\end{pmatrix}
\equiv
\frac{1}{u_1(l;z)}\,\begin{pmatrix}
-u_2(l;z) & 1\\
1 & -u'_1(l;z)
\end{pmatrix}
\end{equation}
It can be directly seen that $s(z)$ is real and self-adjoint for real $z$. Clearly,
the matrix $s$ has simple poles at $\mu_k$, which are at the same time
simple zeros of $u_1(l;z)$. More precisely, by the well-known arguments,
see e.g. Eq.~(I.4.13) in \cite{LS}, there holds
\[
\frac{\partial u_1(l;z)}{\partial z}\Big|_{z=\mu_k}=
u_2(l;\mu_k)\int_0^l u^2_1(s,\mu_k)\,ds,
\]
and $u_2(l;\mu_k)\ne0$ due to $u_1'(l;\mu_k)u_2(l;\mu_k)\equiv w(\mu_k)=1$.
\end{exm}

\section{Reduction to a discrete problem on the lattice}

To describe the spectrum of $\Lambda$ we use the Krein resolvent formula.
Denote by $\Pi$ the operator acting on each edge as $\phi_{m,n,r/u}\mapsto -\phi''_{m,n,r/u}+V\phi_{m,n,r/u}$ on functions satisfying only
the condition \eqref{eq-BC2a1}. Clearly, for such functions the expression $\phi'_{m,n}$ given
by \eqref{eq-BC2b} makes sense.
This operator is not symmetric, as it is
%an
a proper extension of the self-adjoint operator $\Lambda$.

\begin{prop}\label{prop-pgz} The operator $\Pi$ is closed
and the triple $\big(l^2(\ZZ^2),\Gamma,\Gamma'\big)$,
\begin{align*}
\Gamma:\dom \Pi\ni\phi=(\phi_{m,n,r},\phi_{m,n,u})&\mapsto
\big(\phi_{m,n}\big)\in l^2(\ZZ^2),\\
\Gamma':\dom \Pi\ni\phi=(\phi_{m,n,r},\phi_{m,n,u})&\mapsto
\big(\phi'_{m,n}\big)\in l^2(\ZZ^2),
\end{align*}
is a boundary triple for $\Pi$.
\end{prop}

\begin{proof} Denote by $\Xi$ the direct sum of operators $-\dfrac{d^2}{dt^2}+V$
with the domain $H^2[0,l]$ over all edges $E_{m,n,r/u}$. Clearly, $\Xi$ is a closed operator, and
the functionals
\begin{align*}
g_{m,n,1}(\phi)&=\phi_{m,n,r}(0)-\frac{1}{\beta}\,\phi_{m,n,u}(0),\\
g_{m,n,2}(\phi)&=\phi_{m,n,r}(0)-e^{-i\pi n\theta}\phi_{m-1,n,r}(l),\\
g_{m,n,3}(\phi)&=\phi_{m,n,r}(0)-\frac{1}{\beta}\,e^{i\pi m\theta}\phi_{m,n-1,u}(l)
\end{align*}
are continuous with respect to the graph norm of $\Xi$. Therefore,
the restriction of $\Xi$ to functions on which all these
functionals vanish is a closed operator, and this is exactly
$\Pi$. For any $\phi\in\dom\Pi$ the inclusions $\Gamma
\phi,\Gamma'\phi\in l^2(\ZZ^2)$ follow from the Sobolev
inequality, and both $\Gamma$, $\Gamma'$ are continuous with respect
to the graph norm of $\Pi$.

Let $\phi,\psi\in \dom\Pi$, then the integration by parts gives
\begin{multline*}
\langle \phi,\Pi \psi\rangle-\langle\Pi\phi,\psi\rangle\\{}\equiv
\sum_{m,n\in\ZZ,\,i=r,u}\Big(\langle \phi_{m,n,i},-\psi''_{m,n,i}+V\psi_{m,n,i}\rangle-\langle-\phi''_{m,n,i}+V\phi_{m,n,i},\psi_{m,n,i}\rangle\Big)\\{}
\shoveright{\equiv\sum_{m,n\in\ZZ,\,i=r,u}\Big(\langle \phi_{m,n,i},-\psi''_{m,n,i}\rangle-\langle-\phi''_{m,n,i},\psi_{m,n,i}\rangle\Big)}\\
\shoveleft{{}=\sum_{m,n\in\ZZ}\Big\{\overline{\phi_{m,n,r}(0)}\psi'_{m,n,r}(0)+
\dfrac{1}{\beta}\,\overline{\phi_{m,n,u}(0)}\beta\psi'_{m,n,u}(0)}\\
{}-\overline{\phi_{m-1,n,r}(l)}\psi'_{m-1,n,r}(l)
-\dfrac{1}{\beta}\,\overline{\phi_{m,n-1,u}(l)}\beta\psi'_{m,n-1,u}(l)\\
-\overline{\phi'_{m,n,r}(0)}\psi_{m,n,r}(0)
-\dfrac{1}{\beta}\,\overline{\phi'_{m,n,u}(0)}\beta\psi_{m,n,u}(0)\\{}
{}+\overline{\phi'_{m-1,n,r}(l)}\psi_{m-1,n,r}(l)
+\dfrac{1}{\beta}\,\overline{\phi'_{m,n-1,u}(l)}\beta\psi_{m,n-1,u}(l)\Big\}
\end{multline*}
\begin{multline*}
{}=\sum_{m,n\in\ZZ}\Big\{
\overline{\phi_{m,n}}\psi'_{m,n,r}(0)+
\beta\overline{\phi_{m,n}}\psi'_{m,n,u}(0)\\
{}-\overline{e^{i\pi n \theta}\phi_{m,n}}\psi'_{m-1,n,r}(l)
-\beta\overline{e^{-i\pi m \theta}\phi_{m,n}}\psi'_{m,n-1,u}(l)\\
-\overline{\phi'_{m,n,r}(0)}\psi_{m,n}
-\beta\overline{\phi'_{m,n,u}(0)}\psi_{m,n}\\{}
{}+\overline{\phi'_{m-1,n,r}(l)}e^{i\pi n\theta}\psi_{m,n}
+\beta\overline{\phi'_{m,n-1,u}(l)}e^{-i\pi m\theta}\psi_{m,n}\Big\}
\end{multline*}
\begin{multline*}
{}=\sum_{m,n\in\ZZ}\Big\{\overline{\phi_{m,n}}
\big(\psi'_{m,n,r}(0)+\beta\psi'_{m,n,u}(0)\\
{}-e^{-i\pi n \theta}\psi'_{m-1,n,r}(l)- \beta e^{i\pi m\theta}\psi'_{m,n-1,u}(l)
\big)- \big(\overline{\phi'_{m,n,r}(0)}+\beta \overline{\phi'_{m,n,u}(0)}\\
{}-\overline{e^{-i\pi n \theta}\phi'_{m-1,n,r}(l)}- \beta\overline{e^{i\pi m\theta}\phi'_{m,n-1,u}(l)}\big)\,\psi_{m,n}\Big\}\\
{}=\sum_{m,n\in\ZZ} \big( \overline{\phi_{m,n}}\psi'_{m,n}-\overline{\phi'_{m,n}}\psi_{m,n}\big)
\equiv\langle\Gamma\phi,\Gamma'\psi\rangle-\langle\Gamma'\phi,\Gamma\psi\rangle.
\end{multline*}
Now we verify the surjectivity condition. Choose functions
$f_0,f_1\in H^2[0,l]$, with $f_0(0)=f_1'(0)=1$,
$f_0'(0)=f_1(0)=0$, $f^{(j)}_k(l)=0$, $j,k=0,1$.
Let $g,g'\in
l^2(\ZZ^2)$. For any $p,q\in\ZZ$ denote
\begin{align*}
h_{p,q,1}(x)&=g_{p,q}f_0(x)+\dfrac{g'_{p,q}}{4}f_1(x),\\ 
h_{p,q,2}(x)&=\beta g_{p,q}\,f_0(x)+\dfrac{g'_{p,q}}{4\beta}f_1(x),\\
h_{p,q,3}(x)&=g_{p,q}e^{i\pi q\theta}f_0(l-x)+e^{-i\pi q\theta}\dfrac{g'_{p,q}}{4}f_1(l-x),\\
h_{p,q,4}(x)&=\beta g_{p,q}e^{-i\pi q\theta}f_0(l-x)+e^{i\pi p\theta}\dfrac{g'_{p,q}}{4\beta}f_1(l-x),
\end{align*}
then $h_{p,q,j}\in H^2[0,l]$, $j=1,\dots,4$, and these functions satisfy
\begin{gather*}
h_{p,q,1}(0)=\frac{1}{\beta}\,h_{p,q,2}(0)=
e^{-i\pi q\theta}h_{p,q,3}(l)=\frac{1}{\beta}\,e^{i\pi p\theta}h_{p,q,4}(l)=g_{p,q},\\
h'_{p,q,1}(0)=\beta\,h'_{p,q,2}(0)=-e^{i\pi q\theta} h'_{p,q,3}(l)=
-\beta\,e^{-i\pi p\theta} h'_{p,q,4}(l)=\frac{g'_{p,q}}{4},\\
h_{p,q,1}(l)=h_{p,q,2}(l)=h_{p,q,3}(0)=h_{p,q,4}(0)\\
{}=h'_{p,q,1}(l)=h'_{p,q,2}(l)=h'_{p,q,3}(0)=
h'_{p,q,4}(0)=0.
\end{gather*}
Define $\phi^{(p,q)}=\big(\phi^{(p,q)}_{m,n,r},\phi^{(p,q)}_{m,n,u}\big)\in \HH$
with
\begin{gather*}
\phi^{(p,q)}_{p,q,r}=h_{p,q,1},\quad
\phi^{(p,q)}_{p,q,u}=h_{p,q,2},\quad
\phi^{(p,q)}_{p-1,q,r}=h_{p,q,3},\quad
\phi^{(p,q)}_{p,q-1,u}=h_{p,q,4},\\
\phi^{(p,q)}_{m,n,i}=0 \text{ for all other } m,n\in\ZZ \text{ and } i=r,u.
\end{gather*}
Clearly, by construction $\phi^{(p,q)}\in\dom \Pi$ and there holds
$\big(\Gamma \phi^{(p,q)}\big)_{m,n}\equiv
\phi^{(p,q)}_{m,n}=g_{p,q}\delta_{mp}\delta_{nq}$ and
$\big(\Gamma'
\phi^{(p,q)}\big)_{m,n}\equiv\big(\phi^{(p,q)}\big)'_{m,n}=
g'_{p,q}\delta_{mp}\delta_{nq}$, $m,n\in\ZZ$. It is easy to see
that the series $\phi=\sum_{m,n} \phi^{(m,n)}$ converges in the
graph norm of $\Pi$, hence $\phi\in\dom \Pi$. Since $H^2[0,l]$ is
continuously imbedded in $C^1[0,l]$, we have $\Gamma
\phi=\sum_{m,n}\Gamma \phi^{(m,n)}=g$ and $\Gamma'
\phi=\sum_{m,n}\Gamma'\phi^{(m,n)}=g'$. The surjectivity condition
is proved.

It remains to note that the set $\ker\,\Gamma\cap\ker\,\Gamma'$
contains the direct sum of $C_0^\infty(0,l)$ over all edges and
is obviously dense in $\HH$.
\end{proof}

The operator $\Lambda$ is the restriction of $\Pi$ to
the set of function $\phi$ satisfying $\Gamma'\phi=\alpha
\Gamma\phi$. Consider another self-adjoint extension $\Pi_0$ given
by $\Gamma\phi=0$. Clearly, $\Pi_0$ is exactly the direct sum of
the operators $D$ from \eqref{eq-od} over all the edges
$E_{m,n,r/u}$. In particular, $\spec\Pi_0=\spec D$.

Let $z\notin \spec D$, $g\in l^2(\ZZ^2)$
and $\psi_g$ be solution of $(\Pi-z)\psi_g=0$ satisfying the boundary condition $\Gamma\psi_g=g$.
Consider the corresponding Krein function $Q(z):l^2(\ZZ^2)\ni g\mapsto\Gamma'\psi_g\in l^2(\ZZ^2)$.
Application of proposition~\ref{th-krein} gives the following implicit description
of the spectrum of $\Lambda$.
\begin{prop}\label{prop1}
A number $z\in\RR\setminus\spec \Pi_0\equiv\RR\setminus\spec D$
lies in $\spec \Lambda$ iff $ 0\in\spec \big[Q(z)-\alpha\big]$.
Such $z$ is an eigenvalue of $\Lambda$ iff $0$ is an eigenvalue of
$Q(z)-\alpha$.
\end{prop}
Therefore, outside the discrete set $\spec \Pi_0\equiv\spec D$ we can reduce the spectral problem for $\Lambda$ to the spectral problem for $Q(z)-\alpha$. Let us calculate $Q(z)$ more explicitly; actually this is our key construction.
\begin{prop}\label{prop2}
For $z\notin\spec D$ there holds
\begin{equation}
           \label{eq-QM}
Q(z)= (1+\beta^2)\,\big(s_{11}(z)+s_{22}(z)\big)+s_{12}(z) M(\theta,\beta).
\end{equation}
where $M(\theta,\beta)$ is the discrete magnetic Laplacian in $l^2(\ZZ^2)$,
\begin{align*}
\big(M(\theta,\beta)g\big)_{m,n}&=e^{i\pi n \theta}g_{m+1,n}+e^{-i\pi n\theta}g_{m-1,n}
+\beta^2\,(e^{-i\pi m \theta}g_{m,n+1}+e^{i\pi m \theta}g_{m,n-1}),\\
g&=(g_{m,n})\in l^2(\ZZ^2).
\end{align*}
\end{prop}

\begin{proof}
Note that for $\phi=(\phi_{m,n,r},\phi_{m,n,u})$ in the notation of proposition~\ref{prop-pgz}
there holds
\[
\left(\begin{pmatrix}
\phi_{m,n,r}(0)\\
\phi_{m,n,r}(l)\\
\phi_{m,n,u}(0)\\
\phi_{m,n,u}(l)
\end{pmatrix}\right)_{\!\!\!(m,n)\in\ZZ^2}= C \Gamma\phi,\quad
\Gamma'\phi=B\left(\begin{pmatrix}
\phi'_{m,n,r}(0)\\
-\phi'_{m,n,r}(l)\\
\phi'_{m,n,u}(0)\\
-\phi'_{m,n,u}(l)
\end{pmatrix}\right)_{\!\!\!(m,n)\in\ZZ^2}
\]
with operators $B:l^2(\ZZ^2)\otimes\CC^4\to l^2(\ZZ^2)$ and
$C:l^2(\ZZ^2)\to l^2(\ZZ^2)\otimes\CC^4$ given by
\[
B:\left(\begin{pmatrix}
h^{(1)}_{m,n}\\
h^{(2)}_{m,n}\\
h^{(3)}_{m,n}\\
h^{(4)}_{m,n}
\end{pmatrix}\right)\mapsto \big(
h^{(1)}_{m,n}+e^{-i\pi n \theta} h^{(2)}_{m-1,n}+\beta h^{(3)}_{m,n}
+\beta e^{i\pi m \theta} h^{(4)}_{m,n-1}
\big),
\]
and
\[
C:(g_{m,n})\mapsto \left(\begin{pmatrix}
g_{m,n}\\
e^{i\pi n\theta}g_{m+1,n}\\
\beta g_{m,n}\\
\beta e^{-i\pi m \theta}g_{m,n+1}
\end{pmatrix}\right),
\]
cf. \eqref{eq-BC2}.

Let $g\in l^2(\ZZ^2)$. For $z\notin \spec S$, finding the solution $\phi$ with $(\Pi-z)\phi=0$
and $\Gamma\phi=g$ reduces to a series of boundary values problems for components of $\phi$,
\begin{gather*}
\Big(-\dfrac{d^2}{dt^2}+V(t)-z\Big)\phi_{m,n,r/u}(t)=0,\\
\left(\begin{pmatrix}
\phi_{m,n,r}(0)\\
\phi_{m,n,r}(l)\\
\phi_{m,n,u}(0)\\
\phi_{m,n,u}(l)
\end{pmatrix}\right)= C g,\\
\intertext{and}
\begin{pmatrix}
\phi'_{m,n,r}(0)\\
-\phi'_{m,n,r}(l)\\
\phi'_{m,n,u}(0)\\
-\phi'_{m,n,u}(l)
\end{pmatrix}=
\begin{pmatrix}
s_{11}(z) & s_{12}(z) & 0 & 0\\
s_{21}(z) & s_{22}(z) & 0 & 0\\
0 & 0 & s_{11}(z) & s_{12}(z)\\
0 & 0 & s_{21}(z) & s_{22}(z)
\end{pmatrix}
\begin{pmatrix}
\phi_{m,n,r}(0)\\
\phi_{m,n,r}(l)\\
\phi_{m,n,u}(0)\\
\phi_{m,n,u}(l)
\end{pmatrix},
\end{gather*}
For $Q(z)g\equiv\Gamma'\phi$ one has
\begin{gather*}
\Gamma'\psi=B\left(\begin{pmatrix}
\phi'_{m,n,r}(0)\\
-\phi'_{m,n,r}(l)\\
\phi'_{m,n,u}(0)\\
-\phi'_{m,n,u}(l)
\end{pmatrix}\right).
\end{gather*}
Therefore, $Q(z)=B K(z) C$, where $K(z)$
is a linear operator on $l^2(\ZZ^2)\otimes\CC^4$ with the matrix
\[
K(z)=\diag\left(
\begin{pmatrix}
s_{11}(z) & s_{12}(z) & 0 & 0\\
s_{21}(z) & s_{22}(z) & 0 & 0\\
0 & 0 & s_{11}(z) & s_{12}(z)\\
0 & 0 & s_{21}(z) & s_{22}(z)
\end{pmatrix}
\right).
\]
In other words, for any $g\in l^2(\ZZ^2)$ one has
\begin{gather*}
Cg=\left(\begin{pmatrix}
g_{m,n}\\
e^{i\pi n\theta}g_{m+1,n}\\
\beta g_{m,n}\\
\beta e^{-i\pi m \theta}g_{m,n+1}
\end{pmatrix}\right),\\
K(z)Cg=\left(\begin{pmatrix}
s_{11}(z)g_{m,n}+e^{i\pi n \theta} s_{12}(z)g_{m+1,n}\\
s_{21}(z)g_{m,n}+e^{i\pi n \theta} s_{22}(z)g_{m+1,n}\\
\beta\big(s_{11}(z)g_{m,n}+e^{-i\pi m \theta} s_{12}(z)g_{m,n+1}\big)\\
\beta\big(s_{21}(z)g_{m,n}+e^{-i\pi m \theta} s_{22}(z)g_{m,n+1}\big)
\end{pmatrix}\right),
\end{gather*}
and, finally,
\begin{multline}
          \label{eq-Qmn}
\big(Q(z)g\big)_{m,n}=\big(BK(z)Cg\big)_{m,n}\\{}=
(1+\beta^2)\,\big(s_{11}(z)+s_{22}(z)\big)g_{m,n}+e^{i\pi n\theta} s_{12}(z)g_{m+1,n}+
e^{-i\pi n\theta}s_{21}(z)g_{m-1,n}\\{}+ \beta^2 e^{-i\pi m\theta} s_{12}(z)g_{m,n+1}+
\beta^2 e^{i\pi m\theta} s_{21}(z)g_{m,n-1}.
\end{multline}
As can be seen from \eqref{eq-rszA}, there holds $s_{12}(z)=s_{21}(z)$ and
\eqref{eq-Qmn} becomes exactly \eqref{eq-QM}.
\end{proof}

\begin{corol}\label{corol1}
A number $z\in\RR\setminus\spec D$ lies in the spectrum of $L$
iff
\[
0\in \spec \big[(1+\beta^2)\,\big(s_{11}(z)+s_{22}(z)\big)-\alpha+s_{12}(z) M(\theta,\beta)\big].
\]
\end{corol}

\section{Spectral analysis }

To describe the spectrum of $\Lambda$ we need some additional information on the Krein matrix
 $s(z)$ from~\eqref{eq-rszA}.
\begin{prop}\label{prop3} The matrix $s(z)$ has the following properties:
\begin{enumerate}
\item[(A)] $s_{12}(z)\ne 0$ for all $z\notin\spec D$.
\item[(B)] For any $\alpha\in\RR$ the function
\begin{equation}
          \label{eq-etaz}
\eta(z)=\dfrac{\alpha}{s_{12}(z)}-(1+\beta^2)\,\dfrac{s_{11}(z)+s_{22}(z)}{s_{12}(z)}
\end{equation}
can be extended to an entire function. \item[(C)] The function
$\dfrac{1}{1+\beta^2}\,\eta(z)$ is the discriminant of the
(generalized) Sturm-Liouville operator
\begin{equation}
           \label{eq-kp}
P=-\dfrac{d^2}{dt^2}+W(t)+W_{\rm KP}(t),
\end{equation}
where $W$ is the periodic extension of $V$, $W(t+nl)=V(t)$,
$t\in[0,l)$, $n\in\ZZ$, and $W_{\rm KP}$ is the Kronig-Penney
potential $W_{\rm KP}(t)=\displaystyle
\dfrac{\alpha}{1+\beta^2}\,\sum_{k\in\ZZ}\delta(t-kl)$; such
operators $P$ also
%is
are called Kronig-Penney Hamiltonians. \item[(D)] There holds
$\eta(\mu_k)\le-2(1+\beta^2)$ for even $k$ and $\eta(\mu_k)\ge
2(1+\beta^2)$ for odd $k$. \textup{(}Recall that $\mu_k$ are the
eigenvalues of $D$\textup{.)} \item[(E)] For all real $z$ with
$|\eta(z)|<2(1+\beta^2)$ there holds $\eta'(z)\ne 0$. The function
$\eta$ has no local minima with $\eta=2(1+\beta^2)$ and no local
maxima with $\eta=-2(1+\beta^2)$.
\end{enumerate}
\end{prop}

\begin{proof}
Recall that $s(z)$ is given by \eqref{eq-rszA} with $u_1$, $u_2$
from \eqref{eq-uu12}.
%There holds $s_{12}(z)=\dfrac{1}{u_1(l;z)}$.
%The expression $u_1(l;z)$ is an entire function of $z$, so
%$s_{12}(z)\ne 0$ for all $z\notin\spec D$.
There holds $s_{12}(z)=\dfrac{1}{u_1(l;z)}$ and $s_{12}(z)\ne 0$
for all $z\notin\spec D$ since $z\mapsto u_1(l;z)$ is an entire
function. This proves (A).

Substituting \eqref{eq-rszA}  for $z\notin\spec D$ in \eqref{eq-etaz}
one arrives at
\[
\eta(z)=(1+\beta^2)\big(
u_1'(l;z)+u_2(l;z)\big)+\alpha u_1(l;z)
\]
and $\eta$ obviously has analytic extension to all points of $\spec D$. This proves (B).

To understand the meaning of $\eta$ look at the
operator~\eqref{eq-kp}. This operator acts as $f\mapsto-f''+Wf$ on
functions $f\in H^2(\RR\setminus l\ZZ)$ satisfying
\begin{equation}
    \label{eq-dta}
    \begin{pmatrix}
    f'(kl+)\\f(kl+)
    \end{pmatrix}=\begin{pmatrix}
    1 &\dfrac{\alpha}{1+\beta^2}\\
    0 &1
    \end{pmatrix}    \begin{pmatrix}
    f'(kl-)\\f(kl-)
    \end{pmatrix},\quad k\in\ZZ.
\end{equation}
Let $y_1$, $y_2$ be two solutions of $(P-z)y=0$ with $y_1(0+;z)=y'_2(0+;z)=0$ and
$y'_1(0+;z)=y_2(0+;z)=1$. Consider the matrix
\[
M(z)=\begin{pmatrix}
y'_1(l+;z) & y'_2(l+;z)\\
y_1(l+;z) & y_2(l+;z)
\end{pmatrix}.
\]
It is well-known that the spectrum of $P$ consists exactly of real
$z$ satisfying $\tr M(z)\equiv y'_1(l+;z)+
%y'_1(l+;z)\in [-2,2]$,
y_2(l+;z)\in [-2,2]$, see e.g.~\cite{gs,HM}. The function $\tr
M(z)$ is called the discriminant or the Lyapunov function of $P$
and plays an important role in the study of second order
differential operators; if $\alpha=0$, the study of this function
is a classical topic of the theory of ordinary differential
equations, see e.g.~\cite{CL,LS}.

On the other hand, note that on the interval $(0,l)$ the solutions
$y_1$ and $y_2$ coincide with $u_1$ and $u_2$
from~\eqref{eq-uu12}, respectively. In particular,
$y_{1,2}(l-;z)=u_{1,2}(l;z)$ and $y'_{1,2}(l-;z)=u'_{1,2}(l;z)$
Therefore, taking into account the boundary
conditions~\eqref{eq-dta} we can write $M(z)$ in the form
\[
M(z)=\begin{pmatrix}
\dfrac{\alpha}{1+\beta^2}\, u_1(l;z)+u'_1(l;z) & \dfrac{\alpha}{1+\beta^2}\, u_2(l;z)+u'_2(l;z) \\
u_1(l;z) & u_2(l;z)
\end{pmatrix},
\]
and $\tr M(z)=\dfrac{1}{1+\beta^2}\eta(z)$, which proves (C). The items (D) and (E) describe typical properties of the discriminants of one-dimensional periodic operators.

To prove (D) note that for any $k$ one has $u_1(l,\mu_k)=0$, and
$\eta(\mu_k)=(1+\beta^2)\big(\big(u_1'(l;\mu_k)+u_2(l;\mu_k)\big)$, i.e. $\dfrac{1}{1+\beta^2}\eta(\mu_k)$
coincides with the value of the discriminant of the classical periodic Sturm-Liouville problem
($\alpha=0$), for which the requested inequalities are well known, see e.g. Lemma~VIII.3.1 in~\cite{CL}.

The first part of (E) is known for much more general potentials,
see e.g. Lemma~5.2 in~\cite{HM}. As for the second part, local
maxima with $\eta=-2(1+\beta^2)$ and local minima with
$\eta=2(1+\beta^2)$ would be isolated eigenvalues of $P$, which is
impossible, because the spectrum of $P$ is absolutely
continuous~\cite{MS}.
\end{proof}

Therefore, up to the discrete set $\spec D$ the spectrum of
$\Lambda$ is the preimage of $\spec M(\theta,\beta)$ under the
entire function $\eta$. The operator $M(\theta,\beta)$
is very sensitive to the arithmetic properties of $\theta$
and is closely related to the Harper operator, cf.~\cite{Shub}.
The nature of the spectrum of
$M(\theta,\beta)$ in its dependence on $\theta$ is described in
the following proposition, which summarizes theorem 2.7
in~\cite{Shub} (item A), theorem~4.2 in~\cite{Shub}, theorem~1.6
in~\cite{AK} and main theorem in~\cite{10mart} (item B), and
theorem 2.1 in \cite{BZ} (item C).
\begin{prop} \label{prop10}
\begin{enumerate}
\item[(A)] The operator $M(\theta,\beta)$ has no eigenvalues for
all $\theta$~and~$\beta$.
\item[(B)] If $\theta$ is irrational, the spectrum  of
$M(\theta,\beta)$ is a Cantor set. If, in addition, $\beta=1$, the
spectrum has zero Lebesgue measure. \item[(C)] For non-integer
$\theta$ there holds $\|M(\theta,\beta)\| <2(1+\beta^2)$.
\end{enumerate}
\end{prop}

The previous discussion gives a description of
the spectrum of $\Lambda$ in $\RR\setminus\spec D$.
Let us include $\spec D$ into consideration.

\begin{prop}\label{prop5}
There holds $\spec D\subset \spec\Lambda$. Moreover, each $\mu_k\in\spec D$
is an infinitely degenerate eigenvalue of $\Lambda$.
\end{prop}

\begin{proof}

Consider an eigenvalue $\mu_k$ of $D$
and the corresponding eigenfunction $f$ with $f'(0)=1$ and let $\sigma:=f'(l)$.

Let $\theta$ be rational. Take $M\in\ZZ$ such that $\theta M\in2\ZZ$.
Let $p,q\in\ZZ$. Denote by $\phi$ the function from $\HH$ whose only
non-zero components are
\begin{align*}
\phi_{pM+j,qM,r}&=\beta \sigma^jf,& \phi_{pM,qM+j,u}&=-\sigma^jf,\\
\phi_{(p+1)M,qM+j,u}&=\sigma^{M+j}f,& \phi_{pM+j,(q+1)M,r}&=-\beta\sigma^{M+j}f,\\
&&j&=0,\dots,M-1.
\end{align*}
Clearly, $\phi\in\dom\Lambda$ and $-\phi''_{m,n,r/u}
+(V-\mu_k)\phi_{m,n,r/u}=0$ for all $m,n\in\ZZ$. Therefore, $\phi$
is an eigenfunction of $\Lambda$ with the eigenvalue $\mu_k$. As
$p$ and $q$ are arbitrary, one can construct infinitely many
eigenfunctions with non-intersecting supports. Therefore, each
$\mu_k$ is infinitely degenerate in $\spec\Lambda$.

Now let $\theta$ be irrational. We use arguments similar
to the Schnol-type theorems~\cite{Ku2}.
For each $n\in\ZZ$ put
$\phi_{-n,n,r}=\beta e^{i\pi n\theta}f$ and $\phi_{-n,n,u}=-e^{i\pi n\theta}f$.
The ``chain'' constructed from these components does not belong to $\HH$, but
satisfies the boundary conditions \eqref{eq-BC2}. Moreover,  $-\phi''_{n,-n,r/u}+(V-\mu_k)\phi_{n,-n,r/u}=0$ for all $n$,
i.e. this chain is a ``generalized eigenfunction'' of $\Lambda$.

Take $\varphi\in C^\infty[0,l]$ with $\varphi(0)=\varphi'(0)=0$ and
$\varphi(l)=\varphi'(l)=1$.
For any $N\in\NN$ construct $\psi^{(N)}\in\HH$ such that
\begin{gather*}
\psi^{(N)}_{-n,n,r/u}=\phi_{-n,n,r/u} \text{ if } |n|<N,\\
\psi^{(N)}_{-N,N,r}=\varphi \phi_{-N,N,r},\quad
\psi^{(N)}_{N,-N,u}=\varphi \phi_{N,-N,u},\\
\psi^{(N)}_{m,n,i}=0 \text{ for all other } m,n\in\ZZ \text{ and } i=r,u.
\end{gather*}
Clearly, $\psi^{(N)}\in\dom\Lambda$ for any $N$ and $\|\psi^{(N)}\|\ge\sqrt{2N}\|f\|$.
Moreover, the only two non-zero components of
$g^{(N)}=(\Lambda-\mu_k)\psi^{(N)}$ are
\begin{align*}
g^{(N)}_{-N, N,r}&=\beta e^{i\pi N\theta}\big(-\varphi''f-2\varphi'f'-\varphi f''+(V-\mu_k)\varphi f\big)\\
&=-\beta e^{i\pi N\theta}\big(\varphi''f+2\varphi'f\,'\big),\\
\intertext{and}
g^{(N)}_{N, -N,u}&=-e^{i\pi N\theta}\big(-\varphi''f-2\varphi'f'-\varphi f''+(V-\mu_k)\varphi f\big)=e^{i\pi N\theta}\big(\varphi''f+2\varphi'f\,'\big).
\end{align*}
Therefore,
$\|g^{(N)}\|\equiv\|(\Lambda-\mu_k)\psi^{(N)}\|=\sqrt{1+\beta^2}\,
%\big(\varphi''f+2\varphi'f'\big)\equiv C$
\|\varphi''f+2\varphi'f'\|\equiv C$ and
\[
\lim_{N\to\infty}\frac{\big\|(\Lambda-\mu_k)\psi^{(N)}\big\|}{\|\psi^{(N)}\|}\le
\lim_{N\to\infty}\frac{C}{\sqrt{2N}\|f\|}=0,
\]
which means that $\mu_k\in\spec\Lambda$. Let us show that $\mu_k$
is an eigenvalue of $\Lambda$. By proposition~\ref{prop10}(C) one
has $\|M(\theta,\beta)\|<2(1+\beta^2)$. Recall that the spectrum
of $\Lambda$ outside $\spec D$ is the preimage of $\spec
M(\theta,\beta)$ under the function $\eta$ and, due to
proposition~\ref{prop3}(D), does not contain $\mu_k$. As $\mu_k$
is an isolated point of the spectrum, it is an eigenvalue of
$\Lambda$, which is infinitely degenerate according to the
arguments given in remark~\ref{rem-mtg}.
\end{proof}

Now we state the main result of the paper.

\begin{theorem}
The spectrum of $\Lambda$ is the union of two sets,
\[
%\begin{equation}
%    \label{eq-spec}
\spec \Lambda=\Sigma_0\cup\Sigma, \quad \Sigma_0=\spec D,\quad\Sigma=\eta^{-1}\big(\spec M(\theta,\beta)\big),
%\end{equation}
\]
and has the following properties:
\begin{enumerate}
\item[(A)] The discrete spectrum is empty and the point spectrum
coincides with $\Sigma_0$.

\item[(B)] The set $\Sigma$ is non-empty, moreover,
the intersection $[\mu_k,\mu_{k+1}]\cap\Sigma$ is non-empty for any $k$.

\item[(C)] For rational $\theta$ the singular continuous spectrum
of $\Lambda$ is empty and the absolutely continuous spectrum
coincides with $\Sigma$ and has a band structure.

\item[(D)] For irrational $\theta$, the spectrum of $\Lambda$ is
infinitely degenerate. The part $\Sigma$ is a closed nowhere dense
set without isolated points, and $\Sigma\cap(\mu_k,\mu_{k+1})$ is a
Cantor set for any $k=0,1,2,\dots$. If additionally $\beta=1$,
then the spectrum of $\Lambda$ has no absolutely continuous part
and the singular continuous spectrum coincides with $\Sigma$.
\end{enumerate}
\end{theorem}

\begin{proof}

Proposition~\ref{prop5} shows that $\Sigma_0\subset\spec\Lambda$.
The spectrum of $\Lambda$ outside $\spec D$ is described by
corollary~\ref{corol1} and, in virtue of
proposition~\ref{prop3}(B), coincides with $\Sigma$.
%\eqref{eq-spec}.

(A) Propositions~\ref{prop2}, \ref{prop3}(B), and~\ref{prop10}(A)
show the absence of eigenvalues of $Q(z)-\alpha$. In virtue of
proposition~\ref{prop1} the operator $\Lambda$ has no point
spectrum in $\RR\setminus\spec D$ for any $\theta$. Therefore, due
to proposition~\ref{prop5} the point spectrum coincides with
$\spec D$, and all eigenvalues have infinite multiplicity.

(B) The trivial estimate $\|M(\theta,\beta)\|\le 2(1+\beta^2)$ implies the inclusion $\spec M(\theta,\beta)\subset [-2(1+\beta^2),2(1+\beta^2)]$. The assertion follows now from proposition~\ref{prop3}(D).

(C) Let $\theta$ be rational. Take $N\in\ZZ$ with $N\theta\in
2\ZZ$. The operator $\Lambda$ appears to be invariant under the
shifts
$(\phi_{m,n,r},\phi_{m,n,u})\mapsto(\phi_{m+kN,n+lN,r},\phi_{m+kN,n+lN,u})$,
$k,l\in\ZZ$, i.e. is $\ZZ^2$-periodic and, therefore, the absence
of singular spectrum for $\Lambda$ follows from the standard
arguments of the Bloch theory, see e.g. theorem 11 in~\cite{Ku2}.
Therefore by (A) $\Sigma$ coincides with the absolutely continuous
spectrum. The spectrum of $M(\theta,\beta)$ consists of finitely
many bands, so is $\eta^{-1}(\spec M(\theta,\beta))$ between any
two Dirichlet eigenvalues.

(D) Now let $\theta$ be irrational. The infinite degeneracy of
$\spec \Lambda$ follows from the arguments of
remark~\ref{rem-mtg}. The set of $z\in\RR$ for which $|\eta(z)|<
2(1+\beta^2)$ consists of non-intersecting finite intervals $I_n$.
Moreover, due to proposition~\ref{prop3}(D) there is exactly one
such interval between any two subsequent eigenvalues of $D$. Put
$J_n=\overline{I_n}$. Note that $\cup J_n$ contains all points $z$
with $|\eta(z)|\le 2(1+\beta^2)$. Due to
proposition~\ref{prop3}(E), the restriction of $\eta$ to $J_n$ is
a homeomorphism of $J_n$ on the segment
$[-2(1+\beta^2),2(1+\beta^2)]$. Therefore, the preimage
\[
K_n:=\big(\eta|_{J_n}\big)^{-1}\big(\spec
M(\theta,\beta)\big)\subset J_n
\]
is a Cantor set as is true of $\spec M(\theta,\beta)$. Moreover,
the intersection of any two of sets $K_m$ is empty, which follows
from proposition~\ref{prop10}(C) and proposition~\ref{prop3}(D).
Therefore, the set $\cup K_n$, which coincides with $\Sigma$, is
also closed, nowhere dense, and without isolated points.

If $\beta=1$, then the sets $K_n$ are of zero Lebesgue measure due
to proposition~\ref{prop10}(B). Since $(\eta|_{I_n}\big)^{-1}$ are
real analytic, the set $\Sigma=\cup K_n$ is also of zero Lebesgue
measure. Such a set cannot support absolutely continuous spectrum
and does not intersect the point spectrum due to (A), therefore,
$\Sigma$ is the singular continuous spectrum.
\end{proof}

We formulate several corollaries in order to answer the following
natural questions:

\begin{itemize}

\item Are the eigenvalues of $\Lambda$ isolated or embedded in the continuous spectrum?

\item Is the number of gaps in the spectrum finite or infinite?
Note that the rank of the lattice defining the magnetic
translation group is equal to $2$, therefore, one can expect the
validity of the Bethe--Sommerfeld conjecture for $\theta=0$.

\end{itemize}

We emphasize that these questions are rather non-trivial even
for lattices without any potentials; for example, rectangular
lattices with $\delta$-boundary conditions at the nodes can have
very different properties depending on the coupling constants
and the ratio between the edge lengths~\cite{exner-lkp}.
We will see that the introduction of scalar potentials on edges
provides a mechanism of gap creation similar to
the so-called decoration~\cite{AS}.

Let us consider first the case without magnetic field in greater details.

\begin{corol}\label{c2}
Let $\theta$ be integer.
\begin{itemize}
\item[(A)] The part $\Sigma$ of the spectrum of $\Lambda$
coincides with the spectrum of the Kronig-Penney Hamiltonian $P$
from~\eqref{eq-kp}. In particular, if there are inifinitely many gaps
in the spectrum of $P$, then $\Lambda$ has the same property.

%\item[(B)] If $\alpha\ne0$, then the spectrum of $P$ are open,
%and hence of of $\Lambda$ has infinitely many bands.

\item[(B)] If $V$ is a convex smooth function whose derivative
does not vanish, then all gaps are open for any $\alpha$, the
spectrum of $P$ and hence also of $\Lambda$ has infinitely many
gaps, and all $\mu_k$ are isolated in $\spec \Lambda$.

\item[(C)] Let the gap of $P$ near $\mu_k$ be closed for
$\alpha=0$. Then $\mu_k$ is an embedded eigenvalue of $\Lambda$
for all $\alpha$. In particular, $\mu_k$ lies on a band edge for
$\alpha\ne0$. For $V=0$ and $\alpha=0$ all gaps are closed and all
$\mu_k$ are embedded into the continuous spectrum.

\end{itemize}
\end{corol}

\begin{proof}
(A) In this case one has $\spec
M(\theta,\beta)=\big[-2(1+\beta^2),2(1+\beta^2)\big]$ and the set
$\Sigma\equiv\eta^{-1}\Big(\big[-2(1+\beta^2),2(1+\beta^2)\big]\Big)$
coincides with the spectrum of $P$ by proposition~\ref{prop3}(C).

(B)
Denote $\nu(z)=u'_1(l;z)+u_2(l;z)$, where $u_1$ and $u_2$
are the special solutions from example~\ref{ex2}. Clearly,
$\nu$ is the discriminant of the periodic Sturm-Liouville operator
$Q:=-\dfrac{d^2}{dt^2}+W$ with $W$ from~\eqref{eq-kp}.
If $\alpha=0$, then $P=Q$ and $\eta(z)=(1+\beta^2)\nu(z)$ for all $z$.
Let  $V$ be smooth convex with $V'\ne 0$ and $\alpha=0$, then it is
proved in~\cite{gs2} (see lemma~3 and theorem~2 therein) that all
gaps of $P$ are open and that $\mu_k$ do not belong to $\spec
P=\Sigma$, which means that $|\nu(\mu_k)|>2$.
One has $\eta(\mu_k)=(1+\beta^2)\nu(\mu_k)+\alpha
u_1(l;\mu_k)=(1+\beta^2)\nu(\mu_k)$, and the gap remains open
for all $\alpha\ne0$ as $|\eta(\mu_k)|>2(1+\beta^2)$.

(C) The case with $V=0$ and $\alpha=0$ is obvious.
If the gap near $\mu_k$ is closed, then $\nu(\mu_k)=\pm2$,
 $\eta(\mu_k)=\pm2(1+\beta^2)$, and $\mu_k\in\Sigma$. Moreover,
$\nu'(\mu_k)=0$. As $\partial_z u_1(l;\mu_k)\ne0$
(see example~\ref{ex2}), one has $\eta'(\mu_k)\ne 0$
for $\alpha\ne 0$, which means that $\eta\mp2(1+\beta^2)$ changes
the sign at $\mu_k$. This means that
there is a gap near $\mu_k$.
\end{proof}

The case of a non-trivial magnetic field can be treated by a
simple norm estimate.

\begin{corol} If $\theta$ is non-integer, then the spectrum of $\Lambda$
has infinitely many gaps, and all $\mu_k$ lie inside the gap.
\end{corol}

\begin{proof}
In this case the set $\Sigma$ does not contain $\mu_k$ due to
propositions~\ref{prop3}(D) and~\ref{prop10}(D).
\end{proof}

\section{Concluding remarks}

One can easily modify the above proof to many other boundary conditions
at the nodes $K_{m,n}$, for example, for the so-called $\delta'$-boundary conditions,
\begin{equation*}
        \label{eq-BC-kpr}
\begin{gathered}
\big(\frac{d}{dt}-iA_{m,n,r}\big)f_{m,n,r}(0)=\big(\frac{d}{dt}-iA_{m,n,u}\big)f_{m,n,u}(0)\\
{}=-\big(\frac{d}{dt}-iA_{m-1,n,r}\big)f_{m-1,n,r}(l)=-\big(\frac{d}{dt}-iA_{m,n-1,u}\big)f_{m,n-1,u}(l)=:f'_{m,n},\\
f_{m,n,r}(0)+f_{m,n,u}(0)+f_{m-1,n,r}(l)+f_{m,n-1,u}(l)=\alpha f'_{m,n},\quad m,n\in\ZZ,
\end{gathered}
\end{equation*}
A suitable boundary triple would be $(l^2(\ZZ^2),\Gamma_0,\Gamma'_0)$ with
$\Gamma_0=-\Gamma'$ and $\Gamma'_0=\Gamma$ with $\Gamma,\Gamma'$ from proposition~\ref{prop-pgz}.
Instead of the operator $D$ from \eqref{eq-od} one should deal with
the Neumann realization, $Nf=-f''+Vf$, $\dom N=\{f\in H^2[0,l], f'(0)=f'(l)=0\}$,
and the corresponding constructions can be repeated almost literally.

The approach can be extended to the analysis of more general periodic magnetic systems,
for example, for more complicated combinatorial structures,
for nodes and edges with geometric defects or measure potentials, or with
the spin-orbital coupling taken into account.
Such systems will be considered in details in forthcoming publications.

\end{document}